# Using Murphy-Good Plots to Interpret Field Emission Current-Voltage Data

Richard G. Forbes

University of Surrey, Advanced Technology Institute & Dept. of Electrical and Electronic Engineering,
Guildford, Surrey GU2 7XH, UK

*Abstract*—This paper focuses on one small but significant part of a long-term project that aims to improve the interpretation of measured field electron emission (FE) current-voltage [$I_m(V_m)$] data, and (later) the formulation of FE theory. A new form of FE $I_m(V_m)$ data-analysis plot—the so-called "Murphy-Good" (MG) plot"—has recently been introduced, within the general framework of the prevailing "smooth planar metal-like emitter" methodology. This new plot form, and the reasons for its introduction, are discussed. It is shown that the MG plot can perform all the functions that the traditional (90-year old) Fowler-Nordheim plot does, but in addition yields relatively precise results for the characterization parameter "formal emission area". It can be argued that, certainly in scientific contexts, the use of MG plots could usefully replace the use of FN plots.

*Index Terms*—Current-voltage data analysis, field electron emission, Fowler-Nordheim plot, formal emission area, Murphy-Good plot, special mathematical functions, voltage conversion length.

## I. General Introduction

As is well known, field electron emission (FE) is a possible primary cause of electrical breakdown in vacuum, and plays a role in sustaining electrical discharges. It is desirable to have reliable equations to describe FE, both when simulating the possible role of FE in complex discharge phenomena, and when analyzing measured current-voltage [$I_m(V_m)$] data attributed to FE, in order to characterize the emission.

The work described here forms part of a wider long-term project, indicated in another presentation [1] in this conference, and also elsewhere [2], that aims to put FE theory onto a better scientific basis, and eventually to develop greatly improved FE equations. However, for any such improved equations to be effectively tested against experiment, a necessary preliminary is to develop better methodologies for analyzing experimental results, in particular FE $I_m(V_m)$ data. This paper should be seen as a small step towards this long-term aim.

At present, nearly all analysis of measured FE $I_m(V_m)$ data, whatever the emitter material, is carried out within the framework of an underlying basic physical model introduced in the 1920s and used by Fowler & Nordheim [3] in their well-known (but actually faulty [4–6]) 1928 theoretical analysis. This basic physical model uses Sommerfeld-type free-electron theory, disregards the atomic structure of emitters, and assumes that emitters can be modeled as having classically smooth planar surfaces of large lateral extent. The emission is treated as if taking place from a finite area on this surface. I call this *smooth planar metal-like emitter (SPME) methodology*.

Within this SPME methodology, there are alternative choices for the mathematical form of the tunneling barrier (based on different assumptions about tunneling physics). For the last 60 years, FE theoreticians have considered (and still do) that the best *simple* barrier choice in SPME methodology is a planar image-rounded tunneling barrier, often now called a *Schottky-Nordheim (SN) barrier*.

For many modern field emitters (e.g., carbon nanotubes of small apex radius), SPME methodology is obviously far from fully satisfactory, but this is the prevailing paradigm for FE $I_m(V_m)$ data analysis. Some modern FE theories (including some modern simulations related to vacuum breakdown (e.g., [7,8]) do consider non-planar emitters, or do include atomic wave-functions into modeling (e.g., [9,10]), and thus go beyond SPME methodology. These theoretical models are useful in simulations and theoretical explorations, but have not yet been used to develop well-defined and tested methodologies for $I_m(V_m)$ data analysis.

The view of the present author has been as follows. There is an important long-term need to develop better data-analysis and emitter-characterization methods that go beyond SPME methodology, and can also be used to make reliable comparisons between theory and experiment. However, the best short-term option for $I_m(V_m)$ data analysis is to improve SPME methodology. These things can be done in parallel.

Within SPME methodology, the traditional method of current–voltage data analysis has been the Fowler-Nordheim (FN) plot, introduced by Stern et al. in 1929 [4]. (In their 1928 paper (see p. 180), FN could see no possibility of any experimental test of the form of their equation exponent!)

The 1928/29 FN FE equation predicted that a FN plot would be exactly straight. As discussed in [1] (and in more detail in [6]), a conceptual error exists in the FN 1928 paper. A 1928 paper by Nordheim [11] attempted to correct this conceptual error, but, in 1953, Burgess et al. [12] found a mathematical error in the Nordheim paper. A revised theory of FE was then developed by Murphy and Good in 1956 [13]; (for a derivation that uses the modern international equation system, see [14]).

Murphy-Good FE theory predicts that a FN plot will be slightly curved. Thus, for the last 60 years, experimentalists (particularly technologists) have attempted to characterize emitters by fitting a straight line to data points that MG theory predicts to lie on a slight curve. This does not significantly affect the extraction of characterization parameters from the plot slope, but does introduce unhelpful uncertainty into the interpretation of the intercept made on the vertical axis, because the data points typically occupy a limited part of the horizontal axis distant from the vertical axis, and the lever effect operates.

As discussed in [1], there were significant improvements in our mathematical understanding of Murphy-Good theory in the period 2006 to 2010. These included the realization that important information about FE theory may be contained in the plot intercept (or, rather, in the parameter *formal emission area* $A_f^{SN}$ [1] extracted from it—see below), and that it would be useful to improve the accuracy with which values of $A_f^{SN}$ can be extracted from experimental data.

A recent realization is that these new improvements in mathematical understanding do in fact allow the specification of a new plot form, the so-called Murphy-Good (MG) plot [15], that MG theory predicts to be "very nearly straight". This new plot form, although not exactly precise, does take out most of the difficulties associated with FN plots.

The development of this new form, and related theory, was presented in [15]. This poster re-presents the theory and notes some subsequent developments. The distinction between "notional" and "formal" area is discussed here in more detail than in [1] or elsewhere. Universal constants are given to 7 sig. fig., but should be rounded as appropriate.

## II. TECHNICAL INTRODUCTION

To deal with pointed emitters in SPME methodology, one makes the *planar transmission approximation* (see [1]). Then, in zero-temperature MG theory ("MG0"), one can write an expression for the theoretically predicted emission current $I_e^{MG0}$ in the form (see (18) in [1])

$$I_e^{MG0} \equiv A_n^{MG0} J_C^{MG0} = A_n^{MG0} t_F^{-2} J_{kC}^{SN}. \quad (1)$$

Here, $J_C^{MG0}$ denotes the *local emission current density (LECD)* at some characteristic location "C" (in modeling, usually taken as the emitter apex). The *notional emission area (for MG0 theory)*, $A_n^{MG0}$, is defined by (1); $A_n^{MG0}$ will depend on the assumed emitter shape, in a manner not fully known at present. The parameter $t_F$ is the pre-exponential correction parameter derived in 1956 MG theory [13,14].

The *(characteristic) kernel current density for the SN barrier*, $J_{kC}^{SN}$, is defined in terms of the local work function $\phi$ and the local field-magnitude $F_C$ at location "C" by

$$J_{kC}^{SN} \equiv a\phi^{-1} F_C^2 \exp[-v_F b\phi^{3/2}/F_C], \quad (2)$$

where $a$ and $b$ are the *FN constants* (see [1]), and $v_F$ is the appropriate particular value of the *principal field emission special mathematical function* "v", as discussed below. This quantity $J_{kC}^{SN}$ is well defined and can be calculated exactly when $\phi$ and $F_C$ are given.

As discussed in [1], the present author has introduced a slightly modified form of MG0 theory in which the MG pre-factor $t_F^{-2}$ is replaced by a *knowledge uncertainty factor* $\lambda_C$, of unknown functional dependence, and has called this *Extended Murphy-Good (EMG) theory*. Thus, (1) becomes

$$I_e^{EMG} \equiv \lambda_C A_n^{MG0} J_{kC}^{SN} \equiv A_p^{MG0} J_{kC}^{SN}. \quad (3)$$

$\lambda_C$ takes formal account of all physical and theoretical effects neglected in 1956 MG theory, in particular (but not only) the disregard of the influence of atomic wave-functions. The product $\lambda_C A_n^{MG0}$ is then combined into a new formal parameter $A_p^{MG0}$ that (like $\lambda_C$) has unknown value and unknown functional dependence.

As noted in [1], an *ideal FE system* is defined as one in which the measured current and voltage are determined by the system geometry and emission equations alone. An *orthodox FE system* is an ideal FE system for which it can also be assumed that (a) it is adequate to apply Murphy-Good zero-temperature FE theory, and (b) the relevant local work-function value is well known.

For an orthodox system, one can change viewpoint and replace (3) by the "experiment-oriented" equation

$$I_m = A_f^{SN} J_{kC}^{SN}. \quad (4)$$

In this equation, $I_m$ is the directly measurable emission current, and $J_{kC}^{SN}$ can be calculated exactly if $\phi$ and the field $F_C$ are known precisely. Hence, in principle, the *formal emission area (for the SN barrier)* $A_f^{SN}$ is well defined when location "C" is specified (usually as the emitter apex), and is

$$A_f^{SN} \equiv I_m / J_{kC}^{SN}. \quad (5)$$

The label "SN" has to be included because the kernel current densities for different barrier forms are different, but this is the only functional dependence affecting a formal area.

In parallel, in the context of this experiment-oriented approach, the relation between areas in (3) is replaced by

$$A_f^{SN} = \lambda_C A_n^{MG0}. \quad (6)$$

For a theoretician, given an appropriate emitter shape model, and any other necessary information, the parameter $A_n^{MG0}$ can be evaluated precisely. For an experimentalist, as shown below, the parameter $A_f^{SN}$ can now be measured with reasonable precision. However, due to the knowledge uncertainty factor $\lambda_C$ in (3), it is currently impossible to make *precise* scientific comparisons either between these two areas or between emission current density theory and experiment. (And this is likely to remain impossible for many years, due to the intense difficulty of some of the problems apparently involved in determining the values and functional dependences of $\lambda_C$.) The author's view is that *at no stage in the last 90 years* have *precise* scientific comparisons of FE current-density theory and experiment been carried out.

What the above analysis does show, however, is that one can make a separation between: (a) "predictive theoretical activity" (using existing and future FE models); and (b) the activities of experimentalists and of theoreticians who directly support issues of experimental data analysis. The two activities (a) and (b) can thus proceed independently.

It has been shown above that there exists a well-defined parameter ($A_f^{SN}$) that can *in principle* be extracted accurately from experiments. A data-analysis issue is how to do this *precisely*. This paper argues that a Murphy-Good (MG) plot is significantly superior to a Fowler-Nordheim (FN) plot.

For convenience, the problems with FN plots are re-stated next. The underlying problem is that experimentalists are attempting to extract parameters by fitting a straight line to data points that are theoretically predicted (by MG theory) to lie on a curve. Detailed problems are as follows. (1) Three different general extraction methods are in use: (a) modeling the fitted line as a chord to the theoretical FN plot predicted by precise MG theory; (b) modeling the fitted line as a tangent to the theoretical plot; (c) modeling the fitted line by a linearized version of the theory of the theoretical plot. (2) Each different method and approximate formula for $v_F$ yields a different value of $A_f^{SN}$, for the same set of data. (And Russian and US scientists tend to have different preferred formulas). (3) With the tangent method, there are uncertainties related to the issue of how to make the best choice of the "fitting point" at which the fitted line is deemed to be parallel to the tangent to the theoretical plot. (4) Even with the tangent method, a "chord correction" is in principle needed, and its magnitude depends on the precise range (both ends) of the independent variable used to make the plot [17].

The problems with the different approximations for $v_F$ have been eliminated by the modern high-precision formulas for $v_F$ and its derivative (see [1]), but those with the choice of fitting point and the need for a chord correction remain [2].

In reality, the above problems have no significant effect on the accuracy of characterization parameters derived from the FN-plot slope: the problem is to gain improved accuracy and consistency in extracting the formal emission area $A_f^{SN}$. This is where the particular advantage of the MG plot lies.

### III. THEORY OF MURPHY-GOOD PLOTS

*A. Murphy-Good Theory in Expanded Scaled Format.*

As stated in [1], the special mathematical function $v(x)$, where $x$ is the Gauss variable, is applied to MG theory by setting $x$ equal to the *scaled field f* (for a barrier of zero-field height $\phi$) defined in terms of the local surface field $F_L$ by

$$f \equiv c^2 \phi^{-2} F_L = (e^3/4\pi\varepsilon_0)\phi^{-2} F_L, \quad (7)$$

where $c [\equiv (e^3/4\pi\varepsilon_0)^{1/2}]$ is the *Schottky constant*.

As shown in [1], "scaling parameters" can be defined by

$$\theta(\phi) \equiv ac^{-4}\phi^3 \cong (7.433978\times10^{11}\ \text{A m}^{-2})(\phi/\text{eV})^3, \quad (8)$$

$$\eta(\phi) \equiv bc^2\phi^{-1/2} \cong 9.836239\ (\text{eV}/\phi)^{1/2}. \quad (9)$$

For later use, it can be seen that

$$\theta\eta^2 = ab^2\phi^2. \quad (10)$$

By using (8) and (9), the characteristic kernel current density for the SN barrier can then be written in the *scaled format*

$$J_{kC}^{SN} = \theta f_C^2 \exp[-v(f_C)\cdot\eta/f_C]. \quad (11)$$

Using the "simple good approximation" ("F06 formula") for $v(x)$ yields, for characteristic location "C"

$$v_F = v(x=f_C) \approx v_{F06}(x=f_C) = 1 - f_C + (1/6)f_C\ln f_C. \quad (12)$$

Inserting (12) into (11) yields, after some re-arrangement

$$J_{kC}^{SN} = (\theta\exp\eta)\cdot f_C^\kappa \cdot \exp[-\eta/f_C], \quad (13)$$

where the work-function-dependent parameter $\kappa$ is given by

$$\kappa(\phi) = 2 - \eta(\phi)/6 = 2 - 1.639373\ (\text{eV}/\phi)^{1/2}. \quad (14)$$

Equation (13) is an *expanded scaled format* for $J_{kC}^{SN}(f_C)$.

For an ideal FE system, $f_C$ is also "scaled measured voltage", and can be written

$$f_C = V_m/V_{mR}, \quad (15)$$

where the *reference measured voltage* $V_{mR}$ is the $V_m$-value needed to pull the top of the SN barrier down to the Fermi level. Inserting (11) into (10) and (10) into (4) yields

$$I_m = A_f^{SN}(\theta\exp\eta\cdot V_{mR}^{-\kappa})\cdot V_m^\kappa \cdot \exp[-\eta V_{mR}/V_m]. \quad (16)$$

*B. The Form of the Murphy-Good Plot*

A *Murphy-Good plot* has the form $\ln\{I_m/V_m^\kappa\}$ vs $1/V_m$, where $\kappa$ is given specifically by (14). The MG plot is a special case of a more general plot called a "power-$\kappa$" (or "power-$k$") plot. In so-called MG coordinates, (16) becomes

$$\ln\{I_m/V_m^\kappa\} = \ln\{A_f^{SN}\cdot\theta\exp\eta\cdot V_{mR}^{-\kappa}\} - \eta V_{mR}/V_m. \quad (17)$$

Ignoring any weak voltage dependences there may be in $\phi$ or $A_f^{SN}$, the slope $S_{MG}^{th}$ of this theoretical plot is predicted to be

$$S_{MG}^{th}(V_m^{-1}) = d\ln\{I_m/V_m^\kappa\}/d(V_m^{-1}) = -\eta V_{mR}. \quad (18)$$

That is, the slope is predicted to be **constant**.

It follows that the tangent to the theoretical MG plot is predicted to coincide with the theoretical plot, and that this tangent can be defined to meet the vertical axis at $\ln\{I_m/V_m^\kappa\} = \ln\{R_{MG}^{th}\}$, where $R_{MG}^{th}$ is given by

$$R_{MG}^{th} = A_f^{SN}\cdot\theta\exp\eta\cdot V_{mR}^{-\kappa}. \quad (19)$$

*C. Extraction of Slope-related Characterization Parameters*

Now consider that a line with equation

$$\ln\{I_m/V_m^\kappa\} = \ln\{R_{MG}^{fit}\} + S_{MG}^{fit}/V_m \quad (20)$$

has been fitted to an experimental MG plot *taken from an orthodoxly behaving FE system*. On identifying $S_{MG}^{fit}$ with $S_{MG}^{th}(V_m^{-1})$, one concludes from (18) that an extracted value of $V_{mR}$ can be obtained from

$$\{V_{mR}\}^{extr} = -S_{MG}^{fit}/\eta. \quad (21)$$

For any particular measured voltage $V_m$, one can deduce the corresponding characteristic scaled field $f_C$ from

$$\{f_C\}^{extr} = V_m/\{V_{mR}\}^{extr} = -(\eta/S_{MG}^{fit})/V_m^{-1}, \quad (22)$$

and, using (7), the corresponding characteristic field $F_C$ from

$$\{F_C\}^{extr} = \{f_C\}^{extr}\cdot F_R = \{f_C\}^{extr}\cdot c^{-2}\phi^2, \quad (23)$$

where $F_R [=c^{-2}\phi^2]$ is the *reference field* needed to pull the top of the SN barrier down to the Fermi level.

Further, since (for an orthodox system) $V_{mR} = F_R\zeta_C$, where $\zeta_C$ is the system *voltage conversion length* (VCL) (see [1]), it follows from (21) that (a)

$$\{\zeta_C\}^{extr} = -S_{MG}^{fit}/\eta F_R = -S_{MG}^{fit}/b\phi^{3/2}, \quad (24)$$

that (b) $\{\beta_{VC}\}^{extr} = 1/\{\zeta_C\}^{extr}$, and that (c) (when relevant) a *characteristic field enhancement factor (FEF)* $\gamma_C$ is

$$\{\gamma_C\}^{extr} = \underline{d_M}/\{\zeta_C\}^{extr}, \quad (25)$$

where $d_M$ is the *macroscopic-distance parameter* used to define the relevant macroscopic field $F_M$.

All these formulae are free of the tiresome slope correction factor that appears in FN plot theory; also, chord corrections are not needed.

*D. Extraction of Formal Emission Area for SN barrier*

From (18), (19) and (14), it follows that

$$R_{MG}\cdot(|S_{MG}|)^\kappa = A_f^{SN}\theta\exp\eta\cdot\eta^\kappa = A_f^{SN}(\theta\eta^2)\exp\eta\cdot\eta^{-\eta/6}. \quad (26)$$

From earlier, $\theta\eta^2 = ab^2\phi^2$ [$\equiv (7.192492\times10^{-5}\ \text{A nm}^{-2})(\phi/\text{eV})^2$].

Thus, if $S_{MG}^{th}$ and $R_{MG}^{th}$ are identified with $S_{MG}^{fit}$ and $R_{MG}^{fit}$, then an extracted value of formal emission area $A_f^{SN}$ is

$$\{A_f^{SN}\}^{extr} = \Lambda_{MG}\cdot R_{MG}^{fit}\cdot(|S_{MG}^{fit}|)^\kappa, \quad (27)$$

where the *emission area extraction parameter* $\Lambda_{MG}$ *(when using an MG plot that is interpreted using a SN barrier)* is

$$\Lambda_{MG}(\phi) \equiv 1/[(ab^2\phi^2)\cdot\exp\eta\cdot\eta^{-\eta/6}]. \quad (28)$$

Values of $\Lambda_{MG}(\phi)$ are given in [15]. For the specific value $\phi = 4.50$ eV, we have $\eta \approx 4.6369$; $\kappa \approx 1.227$; $\Lambda_{MG} \approx 21.77$ nm$^2$/A. The tiresome factor $(r_t s_t^2)$ has been eliminated from the FN-plot equation [15] for extracted formal emission area.

*E. Applying the Orthodoxy Test Using MG Plots*

Like FN plots, the above MG-plot-based extraction formulas generate valid results only when they are applied to emitters that are behaving in an orthodox fashion. Thus, before attempting to extract characterization parameters, it is advisable to test behavior. The author's "orthodoxy test" ([18], also outlined in [1]) was developed in the context of FN plots, but in fact only requires a reliable method of deducing, from a measured voltage ($V_m$) value, the corresponding value of characteristic scaled field ($f_C$). This can be done using (22) above. So it is clear that the orthodoxy test can be applied to a MG plot. This issue has been explored in a more practical fashion, using simulations, in recent papers [19,20]. As with FN plots, if a FE system fails the test, then rough VCL and FEF estimates can sometimes be found by phenomenological adjustment [21].

### IV. COMPARATIVE TESTING BY SIMULATION

To illustrate the properties of MG plots and the merits of MG plots compared with FN plots, EMG theory (assuming $A_f^{SN}$ constant) was used to create simulated $I_m(V_m)$ data, as below. The two plot types were then used to analyze the data.

For a $\phi$=4.50 eV emitter, a FE system passes the orthodoxy test if the apparent $f_C$-values extracted from a FN or MG plot lies in the range $0.15 \leq f_C \leq 0.45$. There is evidence [17] that most practical tungsten emitters (and, hence, most metal emitters with $\phi$ near 4.5 eV) normally operate within a smaller range inside this. Also, [21] (as interpreted by [22]) suggests that tungsten emitters normally operate within $0.20 \leq f_C \leq 0.35$. In the vacuum breakdown (VB) context, higher operating $f_C$-values may occur. For the simulations discussed here (which were not originally carried out with VB in mind) the range $0.15 \leq f_C \leq 0.35$ was chosen.

Two forms of test have been conducted. The first, partly described in [18], compares FN and MG plot behavior. With input values $\phi$=4.50 eV, $A_f^{SN}$=100 nm$^2$ and $V_{mR}$=8000 V (equivalent to $\zeta_C$=568.87 nm), simulated $I_m$-values were calculated, using (16), (8), (9), the *high-precision* expression for v($x$=$f_C$), and $V_m$=$f_C V_{mR}$. Values were calculated at intervals $\Delta f_C$=0.01. These results were grouped into four sets, as shown in Table 1. For each set, slope values were extracted for both a FN plot and for a SN plot.

Columns 3 and 4 show that (if the physical tunneling barrier is described as a SN barrier, as FE theoreticians believe best practice in SPME methodology [6]), then the slope extraction process is significantly more consistent for an MG plot than for a FN plot. The FN and MG slope values are not expected to be equal: it is the comparative variation that is important: the MG plot behavior is clearly superior. For the MG plot, the residual variations in column 4 presumably result because approximation (12), used to derive MG-plot theory, is not exactly precise.

In the second test, using an MG plot, each of the four data sets is used to provide extracted values of $\zeta_C$ and $A_f^{SN}$. The extraction process is found reasonably consistent, as between the different ranges of scaled field $f_C$. However as compared with the input values, there are small discrepancies, presumably due to the fact that (12) is not exactly precise. For $A_f^{SN}$, the discrepancies are far less than usually expected for FN plots, which can be around 40% or more [2,15].

Note that these tests are intended to assess the self-consistency (with theory) of the MG plot. They are **not** intended as measures of likely accuracy when an MG plot is used to analyze experimental data, with its inevitable errors.

Table 1: Comparison of data extracted from Murphy-Good (MG) and Fowler-Nordheim (FN) plots. (For symbol meanings, see text.)

| $(f_C)_{low}$ | $(f_C)_{up}$ | $S_{FN}^{fit}$ (Np V) | $S_{MG}^{fit}$ (Np V) | $\zeta_C$ (MG) (nm) | $A_f^{SN}$ (MG) (nm$^2$) |
|---|---|---|---|---|---|
| 0.16 | 0.20 | –35902 | –37006 | 567.51 | 99.6 |
| 0.21 | 0.25 | –35577 | –36992 | 567.29 | 98.7 |
| 0.26 | 0.30 | –35256 | –36981 | 567.13 | 98.3 |
| 0.31 | 0.35 | –34938 | –36974 | 567.02 | 97.8 |
| \|Total variation\|: | | 964 | 32 | 0.05 (~0.1%) | 1.8 (~2%) |
| \|Maximum discrepancy\|: | | | | 1.87 (~0.3%) | 2.2 (~2%) |

### V. CONCLUSIONS AND COMMENT

This paper has shown that Murphy-Good plots can be used for all tasks previously done using FN plots, but that MG plots are easier to use and yield more precise numerical data, particularly for formal emission area. This parameter $A_f^{SN}$ may become an important parameter in the development of better FE science. Thus, certainly in science-directed (rather than technology-directed) investigations, MG plots should replace FN plots. (But, actually, MG plots are always better.) The derivations are slightly complicated, but the working extraction formulas, (21)–(25) and (27) are all simple.

Colleagues have argued that multi-parameter numerical fitting might be even better than MG plots. The author's view is that this remains to be convincingly shown, especially if data is noisy, but that this technique is an obvious topic for careful future examination. More prosaically, as a first step it may be easier to persuade FN-plot users to change to using MG plots, rather than to a totally unfamiliar technique.

REFERENCES

[1]  R. G. Forbes, "21st Century planar field emission theory and its role in vacuum breakdown science", oral presentation, this conference.
[2]  R. G. Forbes, J. H. B. Deane, A. G. Kolosko, S. V. Fillipov, & E. O. Popov, 32nd IVNC. & 12th IVESC, Cincinnatti, July 2019. [Technical Digest, p. 23.] doi:10.13140/RG.2.2.32112.81927 .
[3]  R. H. Fowler & L. Nordheim, *Proc. R. Soc. Lond. A* **119**, 173, 1928.
[4]  T. E. Stern, B. S. Gossling & R. H. Fowler, *Proc. R. Soc. Lond. A* **124**, 699, 1929.
[5]  R. G. Forbes & J. H. B. Deane, *Proc. R. Soc. Lond. A* **467**, 2927, 2011. See electronic supplementary material.
[6]  R. G. Forbes, *J. Appl. Phys.* **126**, 210901, 2019.
[7]  A. Kyritsakis & F. Djurabekova, *Comp. Mat. Sci.* **128**, 15, 2017.
[8]  A. Kyritsakis, M. Veske, K. Eimre, V. Zadin & F. Djurabekova, *J. Phys. D. Appl. Phys.* **51**, 225203, 2018.
[9]  B. Lepetit, *J. Appl. Phys.* **122**, 215105, 2017.
[10] C. P. de Castro, T. A. de Assis, R. Rivelino, F de B. Mota, C. N. C. de Castilho & R. G. Forbes, *J. Chem. Inf. Model.* **60**, 714, 2020.
[11] R. E. Burgess, H. Kroemer & J. M. Houston, *Phys. Rev.* **90**, 515, 1953.
[12] L. W. Nordheim, *Proc. R. Soc. Lond. A* **121**, 626, 1928.
[13] E. L. Murphy & R. H. Good, *Phys. Rev.* **102**, 1464, 1956.
[14] R. G. Forbes & J. H. B. Deane, *Proc. R. Lond. A* **463**, 2907, 2007.
[15] R. G. Forbes, *R. Soc. Open Sci.* **6**, 190912, 2019.
[16] R. G. Forbes, & J. H. B. Deane, 30th IVNC, Regensburg, July 2017. [Technical Digest, p. 234.] doi:10.13140/RG.2.2.33297.74083 .
[17] R. G. Forbes, *Proc. R. Soc. Lond. A* **469**, 20130271, 2013.
[18] M. M. Allaham, R. G. Forbes & M. S. Mousa, *Jordan J. Phys.* **12**, 101, 2020.
[19] M. M. Allaham, R. G. Forbes, A. Knápek & M. S. Mousa, *J. Electr. Eng. Slovak* **71**, 37, 2020.
[20] R. G. Forbes, J. H. B. Deane, A. Fischer & M. S. Mousa, *Jordan J. Phys.* **8**, 125, 2015.
[21] W. P. Dyke & J. K. Trolan, *Phys. Rev.* **89**, 799, 1953.
[22] R. G. Forbes, *J. Vac. Sci. Technol. B* **26**, 209, 2010.

E-mail alias of the author: r.forbes@trinity.cantab.net.